\documentclass[aps,prb,floats,twocolumn,superscriptaddress,floatfix]{revtex4}
\usepackage{epsf,graphicx}
\usepackage{amssymb}
\usepackage{amsmath}
\usepackage{eepic}
\usepackage[dvips]{color}
\usepackage{multirow}
\usepackage{color}
\begin{document}

\title{Anisotropic hybridization in a new Kondo lattice compound CeCoInGa$_3$}

\author{Le Wang}
\author{Yuanji Xu}
\author{Meng Yang}
\affiliation{Beijing National Laboratory for Condensed Matter Physics, Institute of Physics, Chinese Academy of Sciences, Beijing 100190, China}
\affiliation{School of Physical Sciences, University of Chinese Academy of Sciences, Beijing 100190, China}
\author{Qianqian Wang}
\affiliation{Beijing National Laboratory for Condensed Matter Physics, Institute of Physics, Chinese Academy of Sciences, Beijing 100190, China}
\author{Cuixiang Wang}
\affiliation{Beijing National Laboratory for Condensed Matter Physics, Institute of Physics, Chinese Academy of Sciences, Beijing 100190, China}
\affiliation{School of Physical Sciences, University of Chinese Academy of Sciences, Beijing 100190, China}
\author{Shanshan Miao}
\author{Youting Song}
\affiliation{Beijing National Laboratory for Condensed Matter Physics, Institute of Physics, Chinese Academy of Sciences, Beijing 100190, China}
\author{Youguo Shi}
\email[]{ygshi@iphy.ac.cn}
\affiliation{Beijing National Laboratory for Condensed Matter Physics, Institute of Physics, Chinese Academy of Sciences, Beijing 100190, China}
\affiliation{School of Physical Sciences, University of Chinese Academy of Sciences, Beijing 100190, China}
\author{Yi-feng Yang}
\email[]{yifeng@iphy.ac.cn}
\affiliation{Beijing National Laboratory for Condensed Matter Physics, Institute of Physics, Chinese Academy of Sciences, Beijing 100190, China}
\affiliation{School of Physical Sciences, University of Chinese Academy of Sciences, Beijing 100190, China}
\affiliation{Collaborative Innovation Center of Quantum Matter, Beijing 100190, China}

\date{\today}

\begin{abstract}
We report a detailed and comparative study of the single crystal CeCoInGa$_3$ in both experiment and theory. Resistivity measurements reveal the typical behavior of Kondo lattice with the onset temperature of coherence, $T^*\approx 50\,$K. The magnetic specific heat can be well fitted using a spin-fluctuation model at low temperatures, yielding a large Sommerfeld coefficient, $\gamma\approx172\,$mJ/mol K$^2$ at 6 K, suggesting that this is a heavy-fermion compound with a pronounced coherence effect. The magnetic susceptibility exhibits a broad field-independent peak at $T_{\chi}$ and shows an obvious anisotropy within the $bc$ plane, reflecting the anisotropy of the coherence effect at high temperatures. These are compared with strongly correlated calculations combining first-principles band structure calculations and dynamical mean-field theory. Our results confirm the onset of coherence at about 50 K and reveal a similar anisotropy in the hybridization gap, pointing to a close connection between the hybridization strength of the low-temperature Fermi-liquid state and the high-temperature coherence effect.
\end{abstract}

\maketitle

\section{INTRODUCTION}
Kondo lattice physics is governed by two competing tendencies towards either coherent heavy-electron state \cite{Kondo1964,Doniach1977,Stewart1984-1,Coleman2001,Yang2016} or long-range magnetic orders of localized $f$-moments \cite{Ruderman1954,Kasuya1956,Yosida1957,Yang2008a}. As a result, it involves a cascade of experimentally well-defined temperature scales, such as the coherence temperature $T^*$, the spin fluctuation temperature $T_{\rm SF}$, and the Fermi liquid temperature $T_{\rm FL}$. Among them, $T^*$ marks the onset of heavy-electron coherence produced by collective hybridization and sets the upper boundary of an intermediate regime with coexisting heavy electrons and unhybridized local $f$-moments \cite{Yang2012}. The transition from fully localized $f$-moments at high temperatures to coherent heavy electrons at low temperatures is at the heart of Kondo lattice physics  \cite{Lonzarich2017}. Consequently, one expects anomalous properties in the intermediate state in all measured quantities, accompanying the emergence of heavy electrons. For example, the susceptibility shows deviation (or even a peak) from its high temperature Curie-Weiss behavior below $T^*$ and the specific heat exhibits logarithmic divergence before it saturates while entering a heavy Fermi liquid ground state. 

Experimentally, these anomalies provide a unified identification of the coherence taking place below $T^*$ \cite{Yang2008a}. However, some also regarded the peak structure in the susceptibility as a way to determine the crystal field scheme. These different opinions reflect the difficulty and confusions in our basic understanding of heavy fermion physics. Moreover, what has been less studied in previous literatures is the anisotropy of the coherence effect \cite{Yang2008b,Ohishi2009} and how this is correlated with the underlying structure of collective hybridization in the Fermi liquid state below $T_{\rm FL}$. Exploration of special anisotropic or even nodal structure of hybridization is becoming a new frontier for novel anomalous and exotic phenomena in heavy fermion research \cite{Dzero2010,Ramires2012,Chen2018}.
 
\begin{figure}[b]
	\includegraphics[width=0.48\textwidth]{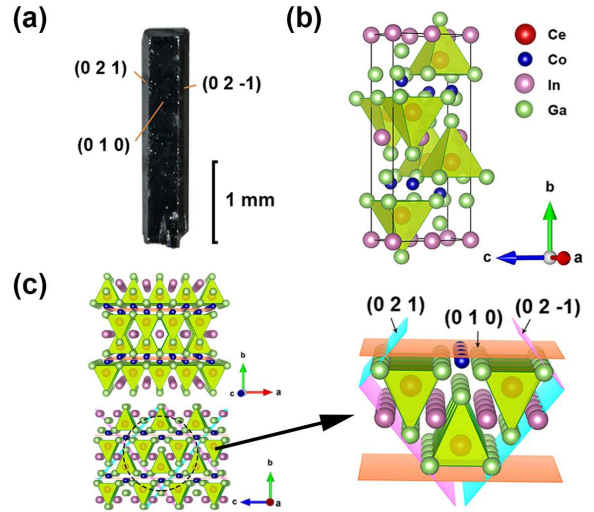}
	\caption{\label{fig1}(Color online) (a) Picture of the CeCoInGa$_3$ single crystal of the size of about 0.5 mm $\times$ 0.3 mm $\times$ 2.5 mm. (b) The orthorhombic unit cell of CeCoInGa$_3$ (space group Cmcm, No. 63). (c) The representation of multiple unit cells. The zone circled by the dashed line is enlarged with the lattice planes indexed as in (a).}
\end{figure}

Here we report the successful synthesis and comparative study of the coherence and hybridization effect in a new Kondo lattice compound CeCoInGa$_3$. Unlike its sister families, Ce-115 such as CeCoIn$_5$ and CeRhIn$_5$ \cite{Petrovic2001,Tayama2002,Park2006,Kenzelmann2008} and Ce-113 such as CeCuGa$_3$ and CeRhGe$_3$ \cite{Hillier2012, Joshi2012,Wang2018}, which show interesting cuprate-like layered structure or noncentrosymmetric structure, respectively, and have thus been widely studied, the Ce-1113 or Ce-114 family has less been investigated, possibly due to the difficulty of its synthesis. In particular, to the best of our knowledge, CeCoGa$_4$ was never studied again ever since its discovery \cite{Routsi1992} where only the magnetic susceptibility was reported in polycrystals to show paramagnetic behavior down to 4.6 K. The 114 family has a quite different structure from 113 and 115, which may be viewed as stacked spin chains of Ce-ions located inside Ga$_5$ pyramids. It is therefore intriguing to see what physics might be discovered in this family. CeCoInGa$_3$ was obtained by doping In atoms into the mother compound CeCoGa$_4$ using the flux method \cite{Canfield1992}. The In atoms occupy the 4a-site of Ga atoms and the crystal structure remains orthorhombic with the space group Cmcm. In doing so, the lattice is expanded and the system is driven further towards a potential quantum critical point, where one may hope to find different quantum critical behavior or even superconductivity. 

We therefore performed a systematic measurement of CeCoInGa$_3$ and our results confirm that it is a standard Kondo lattice compound. The resistivity exhibits a progressive crossover from an insulating-like  state due to incoherent Kondo scattering with logarithmic temperature dependence above $T^*\approx 50\,$K to a Fermi liquid state with $T^2$-dependence below about 6 K. The specific heat exhibits a logarithmic increase due to heavy fermion formation below $T^*$ and contains a $T^3\ln T$ contribution at intermediate temperatures, indicating a possible contribution from spin fluctuations with $T_{\rm SF} \approx 9\,$K. A broad hump is observed in the susceptibility whose position $T_\chi$ varies with the direction of the magnetic field and reflects the anisotropy of the coherence effect. To understand these, we carried out comparative studies using strongly correlated band calculations combining fully consistently the density functional theory and the state-of-the-art dynamical mean-field theory (DFT+DMFT). Our results for CeCoInGa$_3$ produce the correct coherence temperature $T^*$ and reveal an anisotropy of the hybridization gap, which is consistent with the anisotropy in the coherence effect at high temperatures. This suggests that the peak and its anisotropy in the susceptibility may be related to the onset of coherence and its underlying anisotropy of hybridization, which may be further traced back to the formation of Ce-Co-Ce zigzag chains along the $c$-axis rather than the Ce-chains along the shortest $a$-axis.  Unfortunately, we do not find superconductivity down to 2 K in this compound, which indicates that further chemical tuning may be needed for future investigations.

\begin{table}[t]
	\caption{\label{table1}Crystallographic data of CeCoInGa$_3$.}
	\begin{tabular}{ll}		
		\hline
		\hline
		empirical formula & CeCoInGa$_3$ \\
		formula weight & 523.036 g/mol \\
		temperature & 273(2) K \\
		wavelength  & Mo $K_\alpha$ (0.71073 \AA{}) \\
		crystal system & orthorhombic \\
		space group & $Cmcm~(63)$ \\
		unit cell dimensions & $a=4.2315(4)$\AA{} \\
		& $b=16.0755(18)$\AA{} \\
		& $c=6.5974(6)$\AA{} \\ 
		cell volume	& 448.78(8) \AA{}$^3$ \\
		$Z$ & 4 \\
		density, calculated & 7.741 g/cm$^3$ \\
		$h \ k \ l$ range & $-5 \le h \le 5$ \\
		& $-11 \le k \le 20$ \\
		& $-8 \le l \le 7$ \\
		2$\theta_{max}$ & 56.39 \\
		linear absorption coeff. & 36.134 mm$^{-1}$ \\
		absorption correction & multi-scan \\
		no. of reflections & 1184 \\
		$T_{min}/T_{max}$ & 0.004/0.030 \\
		$R_{int}$ & 0.0529 \\
		no. independent reflections & 338 \\
		no. observed reflections & 337 [$F_o > 4\sigma (F_o)$] \\
		$F$(000) & 908 \\
		$R$ values & 5.29 \% ($R_1[F_o > 4\sigma (F_o)]$) \\
		& 13.36 \% (w$R_2$) \\
		weighting scheme & $w=1/[\sigma^2(F_o^2) + (0.0672P)^2$ \\
		& $+ 17.0963P]$, \\
		& where $P = (F_o^2 + 2F_c^2)/3$ \\
		diff. Fourier residues & [-2.969,3.437] e/\AA{}$^3$ \\
		refinement software & SHELXL-2014/7 \\	
		\hline
		\hline	
	\end{tabular}	
\end{table} 

\section{EXPERIMENTAL DETAILS}
Single crystals of CeCoInGa$_3$ were grown using the In-Ga eutectic as flux in alumina crucible sealed in a fully evacuated quartz tube. The crucible was heated to 1100 $^\circ$C for 10 hours and then cooled slowly to 630 $^\circ$C where the flux was spun off by a centrifuge. Rectangle-like single crystals were yielded with the volume of about 0.5 mm $\times$ 0.3 mm $\times$ 2.5 mm as shown in Figure~\ref{fig1}. Elemental analysis was conducted via energy dispersive X-ray (EDX) spectroscopy using a Hitachi S-4800 scanning electron microscope at an accelerating voltage of 15 kV with an accumulation time of 90 s. Single crystal X-ray diffraction was carried out on Bruker D8 Venture diffractometer at 273(2) K using Mo K$\alpha$ radiation ($\lambda=0.71073$ \AA{}). The crystal structure was refined by full-matrix least-squares fitting on $F^2$ using the SHELXL-2014/7 program. A well-crystallized sample was picked out for the measurements. The magnetic susceptibility ($\chi$) was performed in a Quantum Design Magnetic Property Measurement System (MPMS) from 2 K to 300 K under various applied magnetic fields up to 50 kOe in field-cooling (FC) and zero-field-cooling (ZFC) modes. The electrical resistivity ($\rho$) and the specific heat ($C_p$) were measured between 2 K and 300 K in a Physical Property Measurement System (PPMS) using a standard $dc$ four-probe technique and a thermal-relaxation method, respectively. 

\begin{table}[t]
	\caption{\label{table2} Atomic coordinates and equivalent isotropic thermal parameters of CeCoInGa$_3$.}
	\begin{tabular}{lcccccc}
		\hline
		\hline	
		Site  & WP$^a$  & x  & y   & z  & $U_{eq}$  & OP$^b$ \\
		\hline
		Ce  &4c  &0.00000    &0.37989(7)   &0.25000       &0.0160(5)  &1 \\
		Co	&4c  &0.00000   　&0.72432(17)  &0.25000       &0.0163(7)  &1 \\
		In	&4a	 &0.00000	 &0.00000      &0.00000       &0.0239(5)  &1 \\
		Ga1	&4c	 &0.00000	 &0.57916(16)  &0.25000       &0.0202(6)  &1 \\
		Ga2	&8f	 &0.00000	 &0.19182(1)   &0.05409(7)    &0.00173(5) &1 \\
		\hline
		\hline		
	\end{tabular}
\leftline{$^a$Wyckoff position, $^b$Occupation.}
\end{table}	

\section{RESULTS AND DISCUSSION}
The refined results are listed in Tables~\ref{table1} and ~\ref{table2}, indicating a stoichiometric composition with the orthorhombic YNiAl$_4$-type structure (space group Cmcm, No. 63) and the lattice parameters $a=4.2315(4)$ \AA{}, $b=16.0755(18)$ \AA{} and $c=6.5974(6)$ \AA{}. All the crystallographic sites are fully occupied by a unique sort of atoms. The larger In atoms of CeCoInGa$_3$ replace the 4a-site of Ga atoms in CeCoGa$_4$ without changing the crystal structure. It, however, enlarges the inter-plane distance and makes the lattice plane (0 2 1) to be the easy cleavage plane. The Ce atoms locate at 4c site and are each surrounded by five Ga atoms, forming CeGa$_5$ polyhedra that are straightly packed with sharing edges along the $a$-axis. The Co atoms locate between the Ga-cages, forming a layer-like structure. As shown in Fig.~\ref{fig1}(c), the bright crystal surfaces are indexed as (0 2 1), (0 1 0) and (0 2 -1) by single crystal X-ray diffraction, consistent with the enlarged micro-structure. 

\begin{figure}[t]
	\includegraphics[width=0.48\textwidth]{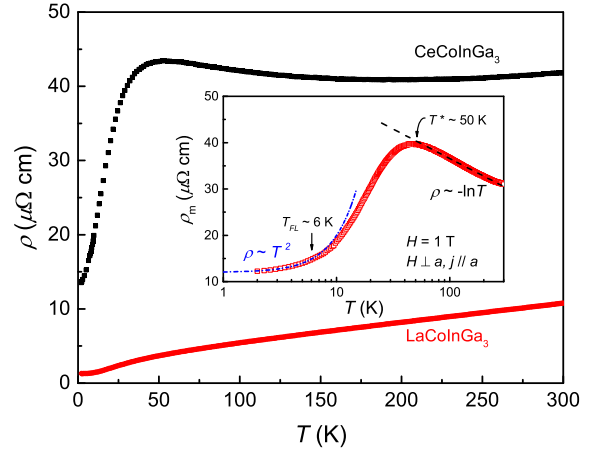}
	\caption{\label{fig2}(Color online) Electrical resistivity of single crystal CeCoInGa$_3$ and LaCoInGa$_3$ with $j \parallel a$, $H = 1$ T and $H \perp a$. The inset shows the temperature dependence of the magnetic resistivity $\rho_m$ after subtracting the resistivity of the isostructural compound LaCoInGa$_3$. A Kondo-type scattering ($\rho_m \sim -\ln T$) was found above the coherence temperature, $T^*\approx50\,$K, as marked by the dashed line. The low-temperature resistivity data can be fitted (dash-dotted line) by the Fermi liquid model, $\rho_m \sim T^2$, giving the Fermi liquid temperature, $T_{\rm FL}\approx 6\,$K. }
\end{figure}

Figure~\ref{fig2} presents the temperature dependence of the $a$-axis resistivity $\rho$ of both CeCoInGa$_3$ and LaCoInGa$_3$ single crystals. The residual resistivity ratio, ${\rm RRR}=\rho(300\,{\rm K})/\rho(2\,{\rm K})$, is 3.1 for CeCoInGa$_3$ and 8.3 for LaCoInGa$_3$. A magnetic field of 1 T perpendicular to the $a$-axis was applied to suppress the superconductivity of the In flux. The magnetoresistance of CeCoInGa$_3$ appears to be very small up to 9 T and is not shown here. The magnetic resistivity $\rho_m(T)$ was obtained by subtracting the nonmagnetic contribution estimated from LaCoInGa$_3$. As shown in the inset of Fig.~\ref{fig2}, it follows a logarithmic temperature dependence above $T^*\approx 50\,$K, indicating a major contribution from Kondo scattering by localized $f$-moments at high temperatures. A larger $T^* \approx 120\,$K has been observed in CeNiAl$_4$ with the same crystal structure \cite{Mizushima1991}, as the lattice is expanded by In and Ga atoms in CeCoInGa$_3$. The coherence peak around $T^*$ in the magnetic resistivity marks the onset of localized-to-itinerant transition. Below 6 K, we find $\rho_m=\rho_0+AT^2$ with a residual resistivity $\rho_0=11.97~\mu\Omega$ cm and a resistivity coefficient $A=0.0826~\mu\Omega$ cm/K$^2$. This defines the Landau-Fermi liquid regime with a Fermi-liquid temperature, $T_{FL}\approx 6\,$K, roughly one tenth of $T^*$ \cite{Kaga1988-1,Kaga1988-2}. We conclude that CeCoInGa$_3$ is a typical Kondo lattice material with a Fermi liquid ground state.

\begin{figure}[t]
	\includegraphics[width=0.48\textwidth]{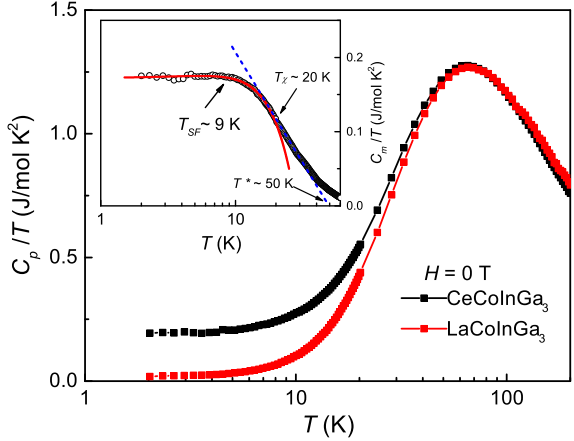}
	\caption{\label{fig3}(Color online) Temperature dependence of the zero-field specific heat coefficient $C_p/T$ for CeCoInGa$_3$ and LaCoInGa$_3$. The inset shows the magnetic contribution $C_m/T$ of CeCoInGa$_3$ after subtracting the lattice contribution estimated from LaCoInGa$_3$. The low-temperature data can be fitted (solid line) using the spin-fluctuation model, $C_m/T=\gamma+DT^2\ln(T_{\rm SF}/T)$ with $T_{\rm SF} \approx 9$ K. The dashed line indicates the logarithmic divergence of the specific heat due to incoherent Kondo scattering above $T^*$. Interestingly, the two lines intersect roughly at $T_{\chi} \approx 20$ K, where a peak is seen in the magnetic susceptibility as shown in Fig.~\ref{fig4}.}
\end{figure}

The specific heat data of CeCoInGa$_3$ and LaCoInGa$_3$ in zero field are compared in Figure~\ref{fig3}, showing no obvious phase transition down to 2 K in both compounds. The magnetic specific heat $C_m$ can be obtained in a similar way by subtracting the lattice contribution estimated from the nonmagnetic LaCoInGa$_3$. As is seen in the inset of Fig.~\ref{fig3}, $C_m/T$ shows a logarithmic divergence with temperature below $T^*$, marking the emergence of heavy electrons accompanying the onset of coherence in the magnetic resistivity \cite{Yang2008a,Yang2016}. Interestingly, at lower temperatures, the magnetic specific heat becomes saturated and  obeys the formula \cite{Trainor1975,Stewart1984-2}, $C_m=\gamma T+DT^3\ln(T_{\rm SF}/T)$, where $T_{\rm SF}$ corresponds to the spin-fluctuation temperature. Our best fit yields the residual specific heat $\gamma=0.172\,$J/mol K$^2$, $D=1.92 \times 10^{-4}\,$J/mol K$^4$, and $T_{\rm SF}\approx 9\,$K. The large $\gamma$ implies a heavy quasi-particle effective mass $m^*$ in the Fermi-liquid state. We can calculate the Kadowaki-Woods ratio, $A/\gamma^2\approx 0.28\times10^{-5}$ $\mu\Omega$ cm (mol K mJ$^{-1}$)$^2$, which is comparable with that of other heavy-fermion compounds \cite{Kadowaki1986}. A rough comparison with the prediction of the spin-1/2 Kondo model suggests a Kondo temperature of about 41 K \cite{Desgranges1982}, roughly consistent with the magnitude of the coherence temperature, but a quantitative fit is impossible. This indicates that in the real system CeCoInGa$_3$ there may be features additionally to the Kondo physics that are not captured by the simple model.

\begin{figure}[t]
	\includegraphics[width=0.48\textwidth]{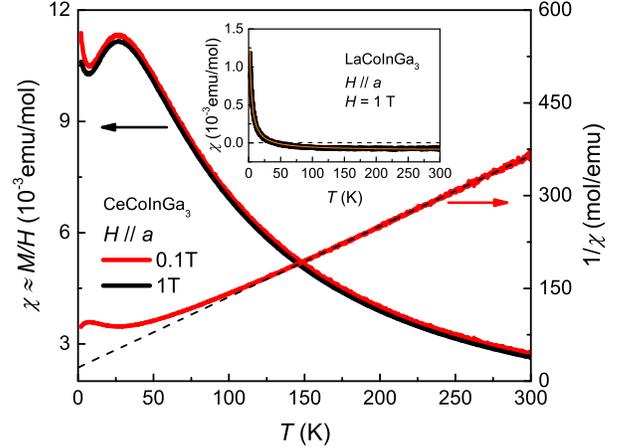}
	\caption{\label{fig4}(Color online) The ZFC susceptibility of CeCoInGa$_3$ with the magnetic field $H=0.1$ and 1 T along the $a$-axis. A best Curie-Weiss fit (dashed line) yields an effective magnetic moment, $\mu_{eff}=2.64\,\mu_B$, and the Weiss temperature, $\theta_p=-19.8$ K. The inset shows the susceptibility of LaCoInGa$_3$ with magnetic field $H = 1$ T along the $a$-axis, revealing diamagnetic behavior at high temperatures. The solid line is a modified Curie-Weiss fit (see text) with $\chi_0=-9.85 \times 10^{-5}$ emu/mol and $\theta_p=-1.35$ K, showing that the Co ions are essentially nonmagnetic.}
\end{figure}

Figure~\ref{fig4} plots the ZFC data of the magnetic susceptibility $\chi$ and the inverse susceptibility $\chi^{-1}$ for $H=0.1$ T and 1 T along the $a$-axis. The $M$-$H$ curve is almost linear up to at least 7 T. A Curie-Weiss fit (dashed line) above 150 K using $\chi(T)=C/(T-\theta_p)$ yields an effective magnetic moment, $\mu_{eff}=2.64~\mu_{\rm B}$, close to the theoretical value of 2.54 $\mu_{\rm B}$ of free Ce$^{3+}$ ion, and a negative Curie temperature, $\theta_p=-19.8$ K. These indicate that the Ce $f$-electrons are well located at high temperatures with an antiferromagnetic exchange coupling. A similar analysis for LaCoInGa$_3$ (inset) using a modified Curie-Weiss formula, $\chi(T)=\chi_0+C/(T-\theta_p)$, yields a diamagnetic background susceptibility $\chi_0=-9.85 \times 10^{5}\,$emu/mol and a Weiss temperature $\theta_p=-1.35$ K, possibly contributed by the Co $3d$-electrons. Thus the Co ions are essentially nonmagnetic. For CeCoInGa$_3$, the violation of the Curie-Weiss behavior below 150 K might be first due to crystal field effects. However, below $T^*$, the development of a broad peak should be attributed to the coherence effect as observed in CeAl$_3$ and URu$_2$Si$_2$ \cite{Yang2012}. In the two-fluid model, it has been argued that increasing hybridization could induce a more rapid delocalization of the localized $f$-moments \cite{Yang2012}. Therefore, the directional dependence of $T_\chi$ potentially reflects the anisotropy of the high temperature coherence effect.

\begin{figure}[t]
\includegraphics[width=0.48\textwidth]{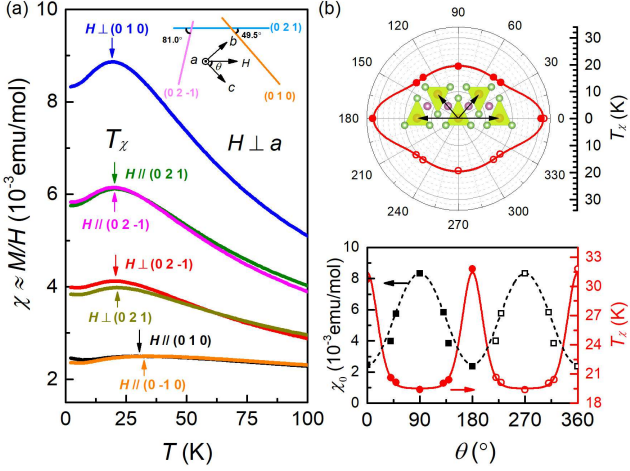}
\caption{\label{fig5}(Color online) (a) The ZFC susceptibility of CeCoInGa$_3$ with magnetic field $H$ perpendicular to the $a$-axis. The angle, $\theta$, is set to zero for $H\parallel c$, as illustrated in the inset. The peak, $T_{\chi}$, marked by the arrows, evolves as the field changes the direction within the $bc$-plane. (b) Angular dependence of $T_{\chi}$ and the residual susceptibility $\chi_0$. The solid circle and square represent the experimental data and the hollow ones are from periodic extrapolation. The upper panel gives the polar diagram of $T_\chi$ in correspondence with the crystal structure. The lower panel compares the angular dependence of $\chi_0$ and $T_{\chi}$.}
\end{figure}

Figure~\ref{fig5}(a) plots the ZFC susceptibilities for field  in parallel with or perpendicular to the (0 1 0), (0 2 1) or (0 2 -1) planes. As shown in Fig.~\ref{fig1}(a), the single crystal of CeCoInGa$_3$ has several facets in one-to-one correspondence with its micro-structure. It is relatively easy to apply the field along these directions. Fig.~\ref{fig5}(b) plots the polar diagram of $T_\chi$ with the data periodically extrapolated to 360$^\circ$. The angle, $\theta$, is set to zero for $H\parallel c$. We see an angular variation associated with the crystal symmetry. Interestingly, as plotted in Fig.~\ref{fig5}(c), there exists an anti-correlation between $T_\chi$ and the residual susceptibility $\chi_0$. For $H \parallel c$, $T_\chi$ is large and $\chi_0$ is small; while for $H \parallel b$, $T_\chi$ is small and $\chi_0$ is large. In the literature, the susceptibility peak has often been attributed to the crystal field effect. We will show that it is potentially correlated with the strength of collective hybridization. Thus the hybridization is stronger along the $c$-axis. In between, the results may be roughly understood by $\chi(\theta)=\chi_c\cos^2\theta +\chi_b\sin^2\theta$.

\section{Numerical calculations}
To gain further insight into above results, we carried out fully consistent DFT+DMFT calculations \cite{Kotliar2006,Haule2010,Held2008}. This method has been successfully applied to CeIrIn$_5$ \cite{Shim2007} and some other materials \cite{Shorikov2015,Yang2007}. However, comparative studies of the hybridization structure on realistic heavy fermion materials are still very few, due to the difficulty in treating the 14 spin and orbital degrees of freedom of the strongly correlated 4$f$-electrons and the extremely low temperature of coherence. For the DFT part, we have adopted the full-potential linearized augmented-plane-wave method as implemented in the WIEN2K package \cite{Blaha2001,Perdew1996}. 

\begin{figure}[t]
\includegraphics[width=0.48\textwidth]{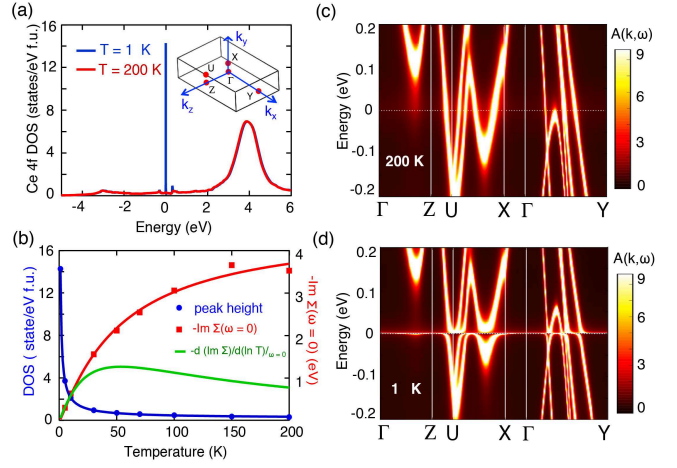}
\caption{\label{fig6}(Color online) (a) Comparison of the Ce 4$f$ density of states (DOS) at 200 K and 1 K calculated using DFT+DMFT. (b) Temperature evolution of the height of quasi-particle peak, the imaginary part of the 4$f$ self-energy at the Fermi energy ($\omega=0$) and its temperature derivative. The results show a maximum at about 50 K and a rapid increase of the DOS at lower temperatures. (c) and (d) compare the momentum-resolved spectral functions at 200 K and 1 K.}
\end{figure}

Figure~\ref{fig6}(a) compares the Ce-4$f$ density of states at 200 K and 1 K using one-crossing approximation as the impurity solver for DMFT \cite{Pruschke1989}. The Coulomb interaction was set to 6 eV, an approximate value typically used for Ce 4$f$-orbitals \cite{Shim2007,Shorikov2015}.The broad peaks at -3 eV and 4 eV correspond to the Hubbard bands. At 1 K, a sharp resonance is seen to develop near the Fermi energy, manifesting the emergence of heavy quasiparticles. This is clearly seen in Fig.~\ref{fig6}(b), where the quasiparticle peak grows rapidly with lowering temperature. Correspondingly, the imaginary part of the self-energy, $|{\rm Im}\Sigma(\omega=0)|$, decreases rapidly, producing a broad maximum at about $T^*\approx50\,$K in its temperature derivative. This is an indication of a crossover in the magnetic scattering rate of the 4$f$-electrons at $T^*$. Above $T^*$, the quasi-particle density of states drops rapidly to zero, marking the loss of heavy electron coherence at higher temperatures.

Figures~\ref{fig6}(c) and \ref{fig6}(d) compare the momentum-resolved spectral function along high symmetry path at 200 K and 1 K. Near the Fermi energy, we see the emergence of evident flat hybridization bands at 1 K, which are not present at 200 K. The hybridization strength may be estimated by fitting each band using $
E_k^\pm=\frac12[ (\epsilon_k+\epsilon_f)\pm \sqrt{(\epsilon_k-\epsilon_f)^2+\Delta^{2}}]$, where $E_k^\pm$ are the two hybridization bands, $\epsilon_k$ is the dispersion of the corresponding conduction band from high temperatures (200 K), $\epsilon_f\approx 0$ is the renormalized $f$-electron energy level, and $\Delta$ corresponds to the direct gap and represents the strength of the hybridization. We obtain $\Delta\approx 22$ meV for the band along the $\Gamma$-Z path ($k_z$), 40 meV along U-X ($k_z$), and 16 meV along U-Z  ($k_y$) and $\Gamma$-Y ($k_x$) paths. Thus the hybridization is stronger along the $c$-axis and weaker along the $a$ and $b$-axes. The origin of such anisotropy might be traced back to the hybridization pathway of the Ce-$f$ electrons. Although one might naively think that the Ce-ions are surrounded by Ga pyramids and connect to form spin chains along the $a$-axis with shortest Ce-Ce distance given by the lattice constant $a$, the Ga-ions seem to play largely the role of a support of the crystal structure, as is the role of B in YbB$_6$ \cite{Zhou2015}, and the hybridization mainly takes place between the Ce-4$f$ and Co-3$d$ bands. The Ce and Co-ions form a zigzag chain along the $c$-axis, favoring the largest hybridization along this direction, while for other two directions, the Ce-Co-Ce bonds are out of plane or have a longer distance, causing their relatively smaller strengths of hybridization. This anisotropy is in good correspondence with the angular variation of $T_{\chi}$, confirming a correlation between the high-temperature coherence effect and the low-temperature hybridization strength.

We would like to further remark that while DFT could sometimes yield useful information for understanding the Fermi surface topology of heavy fermion compounds, it alone cannot describe the development of the $f$-electron coherence with lowering temperature and therefore is incapable of quantitative or even qualitative comparison with many experiments. Moreover, in CeCoInGa$_3$ and many other cases, the Ce-4$f$ bands are predicted in DFT to exhibit a large dispersion near the Fermi energy due to the lack of Kondo renormalization, while the conduction bands are all pushed away. This makes it impossible to derive any information on the hybridization structure between the $f$ and conduction bands. It is only with DFT+DMFT that the $f$-electrons are well treated and strongly renormalized to give rise to flat bands near the Fermi energy, allowing for an unambiguous identification of their hybridization with conduction electrons.

\section{CONCLUSIONS}
We have successfully synthesized high-quality single crystals of CeCoInGa$_3$ and LaCoInGa$_3$ by a flux method. In contrast to it sister Ce-113 and Ce-115 families, the Ce-1113 or Ce-114 family is less well studied. Our systematic investigation of its resistivity, specific heat, and susceptibility provides a unified picture of CeCoInGa$_3$ as a typical paramagnetic Kondo lattice material with logarithmic temperature-dependent specific heat coefficient at low temperatures before the system enters a Fermi liquid state. We identify three important temperature scales in this compound: the coherence temperature $T^*\approx 50\,$K, the spin-fluctuation temperature $T_{\rm SF} \approx 9\,$K and the Fermi liquid temperature $T_{\rm FL}\approx 6\,$K. A broad hump is observed below $T^*$ in the magnetic susceptibility and shows strong anisotropy,  reflecting the directional dependence of heavy-electron coherence. We performed comparative numerical studies. Strongly correlated calculations based on DFT+DMFT confirms the onset of heavy-electron coherence below 50 K, and reveals a similar anisotropy in the hybridization strength, suggesting a close connection with the anisotropy of the coherence effect at high temperatures. We note that replacing Ga by In expands the lattice and drives the system towards a potential quantum critical point where superconductivity may emerge. Although this was not observed in CeCoInGa$_3$, we expect that further chemical tuning will push the system closer to the quantum critical point. A systematic investigation of its peculiar quantum criticality and potential superconductivity, in comparison with the 113 and 115 family, might improve our understanding of heavy fermion physics in association with the crystal structures and hybridization anisotropy.

\section{ACKNOWLEDGMENTS}
This work was supported by the National Key R\&D Program of China (Grant No. 2017YFA0302901, No. 2017YFA0303103, No.  2016YFA0300604), the National Natural Science Foundation of China (NSFC Grant No. 11522435, No. 11474330, No. 11774399, No. 11774401), the State Key Development Program for Basic Research of China (Grant No. 2015CB921300), the Chinese Academy of Sciences (CAS) (Grant No. XDB07020100, No. XDB07020200, No. QYZDB-SSW-SLH043), the National Youth Top-notch Talent Support Program of China, and the Youth Innovation Promotion Association of CAS. 

L.W. and Y.X. contributed equally to this work.

\end{document}